\documentclass[aps,prl,preprint,superscriptaddress]{revtex4-2}
\usepackage{graphicx}
\usepackage{setspace}

\begin{document}


\title{Thickness-dependent  electron momentum relaxation times in thin iron films}


\author{Keno L. Krewer}
	\email{krewer@mpip-mainz.mpg.de}
	\affiliation{
Max Planck Institute for Polymer Research, 55128 Mainz, Germany 
}%
	\affiliation{
Graduate School of Excellence Material Science in Mainz, 55128 Mainz, Germany}%
\author{Wentao Zhang}%
 
	\affiliation{
Max Planck Institute for Polymer Research, 55128 Mainz, Germany 
}%

\author{Jacek Arabski}%
	\affiliation{
Université de Strasbourg, CNRS, Institut de Physique et Chimie des Matériaux de Strasbourg, UMR 7504, 23 rue du Loess, 67034 Strasbourg, France  
}%
\author{Guy Schmerber}%
	\affiliation{
Université de Strasbourg, CNRS, Institut de Physique et Chimie des Matériaux de Strasbourg, UMR 7504, 23 rue du Loess, 67034 Strasbourg, France
}%
\author{Eric Beaurepaire}%
	\affiliation{
Université de Strasbourg, CNRS, Institut de Physique et Chimie des Matériaux de Strasbourg, UMR 7504, 23 rue du Loess, 67034 Strasbourg, France  
}%
\author{Mischa Bonn}

	\affiliation{
Max Planck Institute for Polymer Research, 55128 Mainz, Germany 
}%

\author{Dmitry Turchinovich}

	\affiliation{
Max Planck Institute for Polymer Research, 55128 Mainz, Germany 
}%
	\affiliation{
Fakultät für Physik, Universität Bielefeld, 33615 Bielefeld, Germany}%


\date{\today}

\begin{abstract}
Terahertz time-domain conductivity measurements in 2 to 100~nm thick iron films resolve the femtosecond time delay between applied electric fields and resulting currents. This response time decreases for thinner metal films. The macroscopic response time depends on the mean and the variance of the distribution of microscopic momentum relaxation times of the conducting electrons.  Comparing the recorded response times with DC-conductivities demonstrates increasing variance of the microscopic relaxation times with increasing film thickness. At least two electron species contribute to conduction in bulk with  substantially differing relaxation times.  The different electron species are affected differently by the confinement because they have different mean free paths.
\end{abstract}


\maketitle

Conductivity in metals is typically described using a highly simplified model: a gas of identical electrons characterized by one relaxation time \cite{Fuchs1938,Sondheimer1952,Mayadas1970a,Gall2016},  one mean free path\cite{Fuchs1938,Sondheimer1952,Mayadas1970a,Gall2016,jacob1992,Schad1999}, and hence one velocity.  
This is in stark contrast to the complexity of the underlying process where all electronic states on the Fermi surface contribute to conduction\cite{Abrikosov1988}, and velocity and relaxation time often vary strongly across that large and often complex Fermi surface\cite{Pippard1960,Mott1936}. Therefore, rather than a single value for the relaxation time, one should consider a distribution of relaxation times to describe the system, as has been shown for metal oxides\cite{Kamal2006} and semiconductors\cite{Lloyd-Hughes2012a}.

Scattering of charge carriers is usually discussed in the similarly simplified framework of Mathiessen’s rule, where each scattering process can be assigned a scattering rate independent of any other scattering mechanism present. The scaling of the resistivity is then given by the sum of all rates. For thickness scaling however, the probability of an electron scattering on the surface depends on its mean free path in bulk\cite{Fuchs1938,Thomson1901}. Therefore Mathiessen’s rule should break down and consequently the shape of the relaxation time distribution should change by surface scattering. By time resolving the conduction in iron films we demonstrate this change of the relaxation time distribution, illustrating the breakdown of the simple picture of one relaxation time and Mathiessen’s rule.

We start by deriving the connection between the response time of the macroscopic current $\tau_C$ to the distribution of microscopic momentum relaxation times $\tau$. The exponential decay\cite{Abrikosov1988,Pippard1960} of the conductivity of a state $j$ with a relaxation time of $\tau_j$ is equivalent to a complex conductivity spectrum $\tilde{\sigma}_j (f)$ of Drude shape: 
\begin{equation}
\tilde{\sigma}_j (f)=\underbrace{W_j \tau_j }_{\sigma_{DC,j}}  \frac{1}{1-i2\pi f\tau_j   }  
\label{eq:singleDrude}
\end{equation}

$W_j$, the weight of conduction of state $j$, depends on the group velocity of the state. It connects the relaxation time $\tau_j$ with the steady state conductivity $\sigma_{DC,j}$.  All states $j$ at the Fermi surface will contribute to conduction. We order them by their relaxation time. Thereby, we establish a distribution $w(\tau)$ of relaxation times:
\begin{equation}
w(\tau)=\frac{1}{W} \sum_{\tau_j=\tau}{W_j} \quad  \textnormal{with } W=\sum_{j}{W_j}.  
\label{eq:relaxationTimeDistribution}
\end{equation}
Experimentally, only the total conductivity $\tilde{\sigma}(f)$ is observable. It is the sum of the conductivities of all states $j$ at the Fermi surface. This results in an extended Drude conductivity spectrum\cite{Lloyd-Hughes2012a}. At frequencies $f$  lower than the relaxation rates, we can approximate the extended Drude spectrum as a single effective Drude response:
\begin{equation}
\tilde{\sigma}(f)=\sum_j{\frac{\sigma_{DC,j}}{1-i2\pi f\tau_j }}=  \frac{\sigma_{DC}}{1-i2\pi f\tau_C }+O\left(2 \pi f \tau_j\right)^2    
\label{eq:extendedDrude}
\end{equation}

We can connect the total DC-conductivity $\sigma_{DC}$ and the current response time $\tau_C$ parameterizing this effective Drude response to the distribution of momentum relaxation times.
\begin{equation}
\sigma_{DC}=W\cdot \left\langle \tau \right\rangle; \quad \tau_C=\frac{\left\langle \tau^2 \right\rangle  }{\left\langle\tau \right\rangle  }=\left\langle \tau \right\rangle(1+C^2).
\label{eq:extendedDrudeParameters}
\end{equation}

Here $\left\langle  \right\rangle$ denotes the average of the quantity in brackets over the distribution $w(\tau)$. The coefficient of variation $C$ describes the relative width of the distribution. Contrary to the “ordinary” Drude response of a gas of identical electrons, the ratio $\tau_c/\sigma_{DC}$  is no longer constant for a given material but depends on the variation of the relaxation time distribution. 
We now apply this knowledge to the complex conductivity of thin iron films, measured by terahertz time-domain transmission 
spectroscopy\cite{ulbricht2011} at room temperature  (293~K). The thicknesses $a$ of the iron films range from 2.2 to 100$~$nm. 
The films were deposited on double polished MgO (100) substrates and capped with ca. 12 nm of MgO. The molecular beam epitaxy was performed at room temperature with subsequent annealing. The thicknesses were controlled in situ by quartz balance sensing and confirmed by small-angle x-ray diffraction (XRD) for selected samples (see supplementary material Figure SM1). Roughnesses extracted from XRD average $0.9\pm0.1~$nm, without significant dependence on thickness. Reflection high-energy electron diffraction images indicate that this preparation method achieves single-crystalline films with bcc lattice structure (see Figure SM2 in the supplementary material). More details about the sample preparation and characterisation can be found in the supplementary material.

The terahertz radiation is generated and detected in 1~mm ZnTe crystals using 800~nm 40~fs pulses from an amplified Ti:Sapphire laser emitting 1000 pulses per second\cite{ulbricht2011}. We alternate recording the terahertz transmission through the samples with the transmission through a bare reference substrate. This is repeated 10 to 30 times. We correct the transmission relative to the reference substrate for substrate thickness variations\cite{krewer2018}. We then numerically solve the transfer matrices\cite{Katsidis2002} for the corrected transmission data, using the thin conductive film (‘Tinkham’) approximation\cite{Tinkham1956} to generate starting values.  This approach allows us to reliably determine the phase of the conductivity, even for films for which the complex phase acquired by the terahertz field during a direct transit is non-negligible (see Supplementary Material). 

 \begin{figure}
 \includegraphics[width=0.45\textwidth]{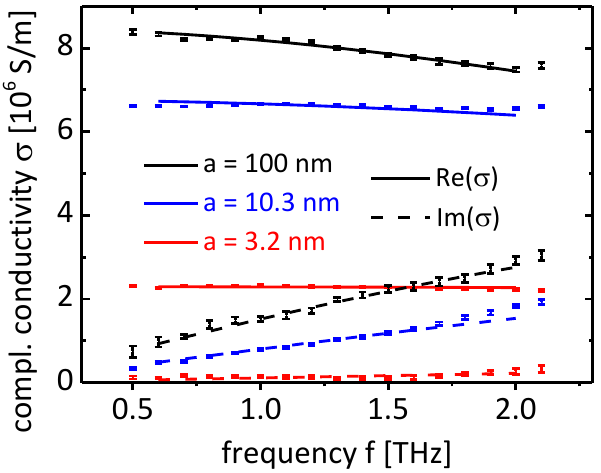}%
 \caption{The complex conductivities $\tilde{\sigma}$ extracted from time-domain spectroscopy for a thick (100~nm, black), intermediate (10.3~nm, blue) and very thin (3~nm, red) sample. Error bars indicate the statistical 68\% confidence interval. Lines denote effective Drude responses. Real conductivity indicated by solid, imaginary by dashed lines.  \label{fig:ConductivitySpectra}}
 \end{figure}

We plot exemplary complex conductivity spectra for a very thin (3.2~nm), an intermediate (10.3~nm) and a thick film (100~nm) in Fig.~\ref{fig:ConductivitySpectra}.  For the 100~nm film, the conductivity has almost converged towards a 'bulk' spectrum. The conductivities are lower for the thinner films, with the difference between 10 to 3.2~nm being much larger than that between 100 and 10~nm. The slope of the imaginary conductivity also becomes much lower for smaller thicknesses.   To quantify these trends, we extract the current response times and DC conductivities of the effective Drude model (eq.~\ref{eq:extendedDrude}). The resulting Drude shapes for the three examples are shown in Fig.~\ref{fig:ConductivitySpectra}.

 \begin{figure}
 \includegraphics[width=0.45\textwidth]{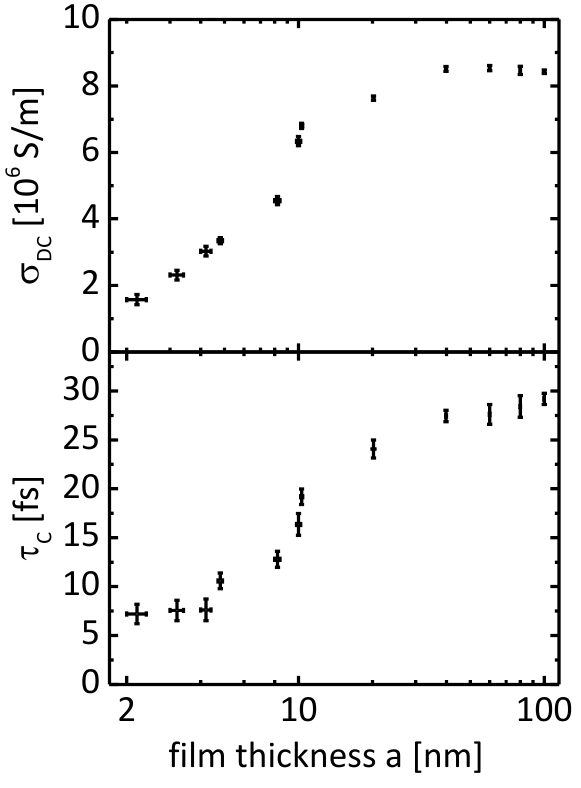}%
 \caption{DC-conductivities $\sigma_{DC}$ (upper) and current response times $\tau_c$ (lower plot) extracted from complex conductivity spectra of 12 iron films from 2.2 to 100 nm thickness. The current relaxation becomes faster for thinner metal films, as expected from increased surface scattering. All measurements were performed at 293 K.   \label{fig:ThicknessDependence}}
 \end{figure}

The extracted DC-conductivities and decay times are shown in Fig. \ref{fig:ThicknessDependence} as a function of sample thickness. The conductivities decrease with decreasing thickness. We attribute the reduction of conductivity mainly to surface scattering\cite{Fuchs1938,Sondheimer1952}. Other effects, such as increased defects and relative film roughness with decreasing thickness\cite{Namba1970} may also play a minor role. The response time, determined here with $\approx$1~fs precision, decreases monotonously with decreasing thickness. Previous measurements of current response times as function of metal thicknesses yielded constant results within their error of 10~fs for polycrystalline gold films\cite{Walther2007}. Here, we directly observe increased scattering by constriction in a metal.

	 \begin{figure*}
 \includegraphics[width=1\textwidth]{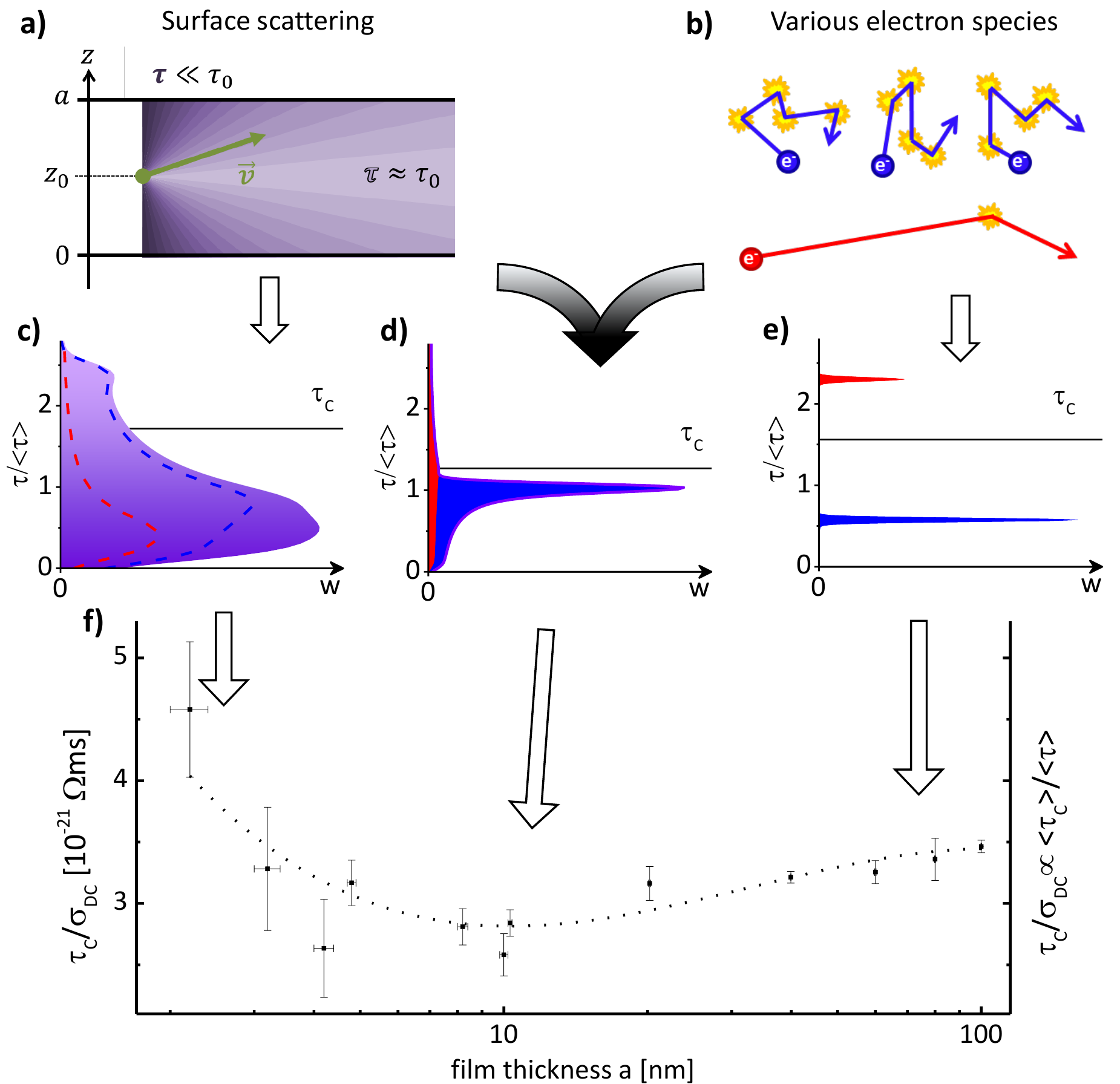}%
 \caption{  a) Effect of surface scattering: variation in and a decrease of momentum relaxation times compared to bulk. The relaxation time depends on position and velocity of an electron in a thin film. b) Illustration of 2 carrier species with different momentum relaxation. The blue species is more numerous, but scatters more often than the red one. The red species has hence a longer mean free path. c-e) Sketched distributions of relaxation times due to different carrier species and interface scattering. The relaxation times are displayed relative to the mean relaxation time; The weight densities $w$ are scaled to their respective maxima.  c) Shows the distribution for an extremely thin film, where interface scattering dominates for both species. d) Distribution for a thickness lower than the mean free path of species “red” but higher than that of “blue”. e) Distribution of 2 carrier species as sketched in b).  f) Measured ratio between current decay time and dc-conductivity as a function of thickness. The dotted line is a 3rd order polynomial fit to guide the eye.    \label{fig:distributionChange}}
 \end{figure*}

If all electrons conducting in iron had the same relaxation time, this relaxation time would be identical to the response time and directly proportional to the measured DC-conductivity (see eq. \ref{eq:singleDrude} and \ref{eq:extendedDrudeParameters}). However, while both response time and conductivity decrease, the response time decreases faster than the conductivity for film thicknesses above 10~nm, and slower below. We investigate this discrepancy by calculating the ratio between the current response times and the conductivities, shown in fig. \ref{fig:distributionChange} f). This ratio is large for very thin and very thick films, with a minimum around 10~nm.

This ratio $\tau_c/\sigma_{DC} =((1+C^2 ))/W$ depends on the coefficient of variation $C$ of the distribution of the relaxation times; the wider the distribution, the larger the ratio. This means a wide relaxation time distribution exists in bulk, the distribution is squeezed together at intermediate thicknesses and then broadens again towards the thinnest films. Surface scattering broadens the relaxation time distribution since the relaxation time of each electron will now depend on its distance to and velocity towards the surface (sketched in Fig. \ref{fig:distributionChange} a). The large spread of relaxation times towards the bulk proves that there are electron species with different relaxation times conducting in bulk iron. We illustrate how surface scattering may squeeze the relaxation times of different species together for intermediate thicknesses by considering the simple case of two electron species conducting in bulk, “blue” and “red” (Fig. \ref{fig:distributionChange} b). Species “blue” has a shorter relaxation time and a lower velocity, but is more numerous than its “red” counterpart.  The relaxation time of each individual species has no variance, but combined into a single distribution, the large difference between the two relaxation times translates to a large variance; the combined distribution (Fig. \ref{fig:distributionChange} e) is spread out widely.  Surface scattering reduces the relaxation times and broadens the distribution of each individual species.  Since the “red” electrons have a much longer mean free path, surface scattering will affect them already at larger film thicknesses compared to the “blues”. Hence at intermediate thicknesses, surface scattering shifts the “red” distribution to lower relaxation times and broadens it, while the “blue” distribution remains largely unchanged. The “red” distribution moves towards the “blue” and the two overlap. This overlap of the two species overcompensates the increased variance of the individual species. Therefore the combined distribution is less spread out. This leads to a reduction in the coefficient of variation like we observe between 100 and 10~nm. When the thickness becomes smaller than the “blue” mean free path (Fig. \ref{fig:distributionChange} c), interface scattering dominates for both species, resulting in a very broad combined distribution. This is consistent with the large $\tau_c/\sigma_{DC}$  ratio at 2.2~nm, however we note that the statistical significance of the change in ratios is much smaller for the thinnest films compared to the intermediate and thick films.  Averaging over the variety of thicknesses in a rough film will decrease the observed conductivity\cite{Namba1970} more than the observed response time. This systematic distortion of the $\tau_c/\sigma_{DC}$  ratio will only appear in the thinnest films. 

Therefore we focus on our analyses of the wide spread of relaxation times in bulk. We estimate the minimum spread in bulk by comparing the ratios between the 10.3~nm and 100~nm samples. The minimal possible coefficient of variation C is 0, which we assume for 10.3~nm. The ratio $\tau_c/\sigma_{DC}$ at 100~nm is 22\% larger than that at 10.3~nm. Using eq. (4), this yields a minimum coefficient of variation of 46\% at 100~nm. We use this assumption to estimate parameters of the relaxation time distribution. The average relaxation time $\left\langle \tau \right\rangle$ must be less than 82\% (24~fs) of the current response time (29~fs) in this sample. We note that the conductivity at 100 nm is only 84\% of optimally annealed bulk iron\cite{Hust1984}, offsetting this correction for a maximum average relaxation time in bulk. The full width (double standard deviation) of the relaxation time distribution is 22~fs . The average Fermi velocity of iron is ca. 0.2 nm/fs\cite{Schafer2005}, from which we estimate a maximum average mean free path of  6~nm. This is a factor of 2 lower than estimates\cite{jacob1992,Schad1999} based on a comparison of DC-resistivity measurements and Fuchs-Sondheimer\cite{Fuchs1938,Sondheimer1952} theory. If our assumption of surface scattering driving the change in the relaxation time distribution holds, the relative variation in the bulk mean free paths will be similar to that deduced for the relaxation times. This variation of mean free paths may explain the discrepancy in mean free path estimates. 

Iron is a ferromagnet; its band structure is spin split. Since Mott\cite{Mott1936}, a spin splitting of the relaxation times has been accepted as cause for its magneto resistance.  However, iron has 4 different bands\cite{Schafer2005} per spin at the Fermi surface. This means that one cannot infer that the observed variation results only from the difference between the relaxation times and mean free paths for charge carriers with opposite spin.  Spin split relaxation times have been observed in multiple ferromagnetic layers arranged as spin valves\cite{jin2015}. However, a debate remains whether that split is predominantly caused by scattering inside the layers or at their interfaces. The intrinsic spread we observe here would be large enough to cause the observed giant magnetoresistance effect if much of the spread indeed resulted from spin splitting. 

We also note that the relative contribution to conduction of the individual species changes when the surface scattering increases depending on their respective mean free paths. Revisiting our simplistic model of a “red” and a “blue”  species, species “red” carries a majority of current in bulk but only a small minority in the surface scattering limited case. This means the “color” polarization of the current changes in magnitude and sign. If “color” correlates with a property such as spin or spin orbit coupling, this will affect the thickness scaling relations of any effect resulting from that property (i.e. the anomalous/spin Hall effect\cite{Hou2015}). 

We have shown that terahertz time-domain spectroscopy is capable of resolving current response times $\tau_c$ with $\approx$1~fs accuracy in metal films where these response times are on the order of 10~fs. The current response time $\tau_c$ is different from the average relaxation time $\left\langle \tau \right\rangle$ by a factor depending on the spread of the relaxation times. The ratio between response time and conductivity $\tau_c/\sigma_{DC}$  allows to quantify the spread of relaxation times as a function of thickness for our iron films. The thickness scaling relation of this ratio proves that a wide distribution of electron relaxation times exists in bulk iron and that the relative width and therefore the shape of this distribution depends on the film thickness. This demonstrates that surface scattering does not obey Mathiessen’s rule. 

We are grateful to Eduard Unger for automatizing the measurements and to Zoltan Mics for building the high precision set-up used.  
K.K. acknowledges the support from MAINZ - Graduate School of Excellence Material Science in Mainz. D.T. acknowledges the project “Nonequilibrium dynamics in solids probed by terahertz fields” funded by the Deutsche Forschungsgemeinschaft (DFG, German Research Foundation) – Projektnummer 278162697 – SFB 1242.

\bibliography{thicknessDependentRelaxationTime}

\end{document}